\documentclass{article}
\usepackage{float}
\usepackage{amsfonts}
\usepackage{graphicx}
\usepackage{amsmath}
\usepackage{appendix}
\usepackage{multicol}
\usepackage{fullpage}
\usepackage{hyperref}
\usepackage{cite}
\usepackage{setspace}
\hypersetup{
  colorlinks   = true, 
  urlcolor     = blue, 
  linkcolor    = blue, 
  citecolor    = blue  
}

\begin{document}

\title{\huge{ \textbf{Macroscopic phase resetting-curves determine oscillatory coherence and signal transfer in inter-coupled neural circuits.}}}

\date{}

\author{Gr\'{e}gory Dumont\footnote{\'{E}cole Normale Sup\'{e}rieure,
Group for Neural Theory, Paris, France, email: gregory.dumont@ens.fr},
          Boris Gutkin \footnote{\'{E}cole Normale Sup\'{e}rieure, Group for Neural Theory, Paris, France, email: boris.gutkin@ens.fr}
      }

\maketitle

\section*{Abstract}
Macroscopic oscillations of different brain regions show multiple phase relationships that are persistent across time and have been implicated routing  information.  Various cellular level mechanisms influence the network dynamics and structure the macroscopic firing patterns. Key question is to identify the biophysical neuronal and synaptic properties that permit such motifs to arise and how the different coherence states determine the communication between the neural circuits. We analyse  the emergence of phase locking within bidirectionally delayed-coupled spiking circuits showing global gamma band oscillations.  We consider both the interneuronal (ING) and the pyramidal-interneuronal (PING) gamma rhythms and the inter coupling targeting the pyramidal or the inhibitory interneurons. Using a mean-field approach together with an exact reduction method, we break down each spiking network into a low dimensional nonlinear system and derive the macroscopic phase resetting-curves (mPRCs) that determine how the phase of the global oscillation responds to incoming perturbations.  Depending on the type of gamma oscillation, we show that incoming excitatory inputs can either only speed up the oscillation (phase advance; type I PRC) or induce both an advance and a delay the macroscopic oscillation (phase delay; type II PRC). From there we determine the structure of macroscopic coherence states (phase locking) of two weakly synaptically-coupled networks. To do so we derive a phase equation for the coupled system which links the synaptic mechanisms to the coherence state of the system.  We show that the transmission delay is a necessary condition for symmetry breaking, i.e. a non-symmetric phase lag between the macroscopic oscillations, potentially giving an explanation to the experimentally observed variety of gamma phase-locking modes.  Our analysis further shows that symmetry-broken coherence states can lead to a preferred direction of signal transfer between the oscillatory networks and how this depends on the timing of the signal. Hence we propose a theory for oscillatory modulation of functional connectivity between cortical circuits.


\section*{Author summary}
Large scale brain oscillations emerge from synaptic interactions within neuronal circuits. Over the past years, such macroscopic rhythms have been suggested to play a crucial role in routing the flow of information across cortical regions, resulting in a functional connectome.  The underlying mechanism are cortical oscillations that bind together following a well-known motif called phase locking. While there is significant experimental support for multiple phase-locking modes int eh brain, it is still unclear what is the underlying mechanism that permits macroscopic rhythms to phase lock. In the present paper we take up with this issue, and to show that, one can formulate emergent macroscopic phase-locking within the mathematical framework of weakly coupled oscillators. 
We thus offer clarification on the synaptic and circuit properties responsible for the emergence of multiple phase locking patterns and provide support for its functional implication in information transfer.

\section*{Introduction}

Ranging from infraslow to ultrafast, brain rhythms is a nearly omni-present phenomenon covering more than four orders of magnitude in frequency. Of this variety of rhythms, the class of gamma oscillations falling in the specific frequency band of $30-150$ Hz - is arguably the most studied rhythmic activity pattern \cite{Buzsaki2004,buzsaki2006rhythms}. It has been reported in many brain regions, across many species, and is associated with a variety of cognitive tasks  \cite{buzsaki2013,Fries2007}. There is nowadays growing evidence that the gamma cycle results from emergent dynamics of cortical networks, a natural consequence of the interplay between interconnected pyramidal cells and subnetworks of interneurons \cite{Wang2012,Bartos2007}.

Although brain rhythms such as gamma rhythms emerge locally  \cite{Bartos2007}, they are known to interact in a coherent fashion across the cortical scale \cite{Fries2005,Bartos2017}. As such, macroscopic oscillations within different brain regions show multiplt phase relationships that are persistent across time \cite{fries2009neuronal}. Crucial for a prominent theory of how oscillations shape the information transfer within and across the cortex, the communication through coherence (CTC) hypothesis, such cross-coupling is believed to be implicated in a number of higher cognitive functions. 
For example, enhanced interareal gamma-band coherence is  considered as the neural correlate of selective attention, in which a network receiving several informational stimulus can preferentially react to one or another depending on task relevance \cite{Fries2005}.

The CTC hypothesis provides a mechanism by which gamma rhythms can participate in regulating the information flow \cite{buzsaki2006rhythms}. The rationale behind is that gamma oscillations are the consequence of rhythmic inhibitory feedback inducing an hyperpolarization of the principle cell membrane potential \cite{Wang2012,Bartos2007}. Synaptic inputs targeting excitatory cells are then expected to cause a stronger reaction when the inhibition drops off. This gives rise to a temporal window of excitability within the oscillatory cycle during which pyramidal neurons are more likely to respond to stimulation \cite{Cardin2009}. Ongoing oscillatory firing patterns rhythmically modulate the excitability of networks, and therefore, two neural groups engaged in a rhythmic dynamic communicate more efficiently when they maintain a coherent relationship: they can consecutively send their information at the most excitable phase \cite{Fries2005,Fries2007}.

According to the CTC hypothesis, neuronal interactions and transfer of information are dynamically shaped by the phase relationship between neuronal oscillations \cite{womelsdorf2007modulation}. In fact it has been proposed that macroscopic rhythms offer a way of adjusting the effectivity of functional connectivity while leaving untouched the anatomical connections \cite{fries2009neuronal} and resulting in a functional connectivity  \cite{Battaglia2012,Palmigiano2017}. This functional connectivity, often defined in correlational or information transmission terms, is determined by the relative phase relationship between the communicating networks. Note that an optimal locking mode does not always result from a zero phase lag or perfect synchrony. The reason is that, spike transmission from one network to another is not instantaneous and, depending on the distance, projection across the brain can take up to hundreds of milliseconds \cite{delay}. Therefore oscillations should be lagged in order to see their spikes arriving at the most excitable phase. This most excitable phase also depends on the biophysical properties of the constituent neurons and of the emergent rhythms (e.g. as characterised by the network-wide phase response curves \cite{Dumont2017}. An optimal phase difference will thus depend on the properties of the neural groups at work and the distance between the two \cite{Deco2016,Fries2016}. Recent experimental studies have reported a multiplicity a phase differences and it has been argued that such a diversity might facilitate information selectivity \cite{Fries2016}.

Over the past few years, computational studies have devoted a great deal of attention to uncovering the precise functional roles of gamma patterns and gamma interaction. Doing so, they have been able to reproduce experimental findings in support of several predictions of the CTC hypothesis. For instance, modeling approaches have shown that the gamma cycle generates a temporal window of excitability \cite{Knoblich2010}, which is suitable to suppress irrelevant stimuli \cite{Borgers2007,Borgers18112008}. Others studies have demonstrated that the mutual information between two neural groups engaged in rhythmic patterns is tuned with respect to their phase lag \cite{Deco2010, Barardi2014}, and a directionality in the flow of information emerges through a symmetry breaking in the phase relationship \cite{Battaglia2012,Palmigiano2017}. A diversity of phase lags can then be observed which benefits information coding and stimulus reconstruction \cite{Lowet2015}. Finally, in a rather different line of thinking from the main current view of CTC, computational studies have exposed how cortical oscillations could implement a multiplexing \cite{Akam2010,Akam2012a,Akam2014}.

However, the underlying mechanisms responsible for the emergence of the multiple phase-locking modes and of the ensuing functional connectivity as proposed by the CTC are not trivial. So far, no mechanistic view to explain the observed variety of phase lags has been proposed. The question is then to identify through what synaptic mechanisms can these rhythms coordinate their temporal relationships in such a diversity of locking modes. Answering this question is crucial and knowing the chain of causation that allows for coherent oscillations is key to understanding their functional role \cite{Canavier2015, kopell2014}. Hence, an subsequent question is how can one characterize the functional connectivity associated with the various phase locking modes and how directed signal transmission can ensue.

We investigate the dynamical emergence of phase locking within two bidirectionally delayed-coupled spiking networks. Importantly, the neurons within the circuits have a relatively wide distribution of intrinsic excitability, meaning that most of them not intrinsically oscillating. Hence the gamma rhythm in our network is an emergent property of the global dynamics, as opposed to phasing of coupled oscillators (see \ref{STROGATZ20001} for instance). Furthermore, the design of the interconnections between our networks is inspired from previous research  \cite{Deco2010, Barardi2014,Palmigiano2017}  to essentially capture multiple communicating brain regions where transfer of information takes place.
Each network is assumed to be made up of pyramidal cells and interneurons, and each cell is characterized by a conductance-based neural model \cite{Izi, Ermentrout}.  A synaptic delay is included to account for possible long range distances separating the circuits  \cite{delay}. We then take advantage of a thermodynamic approach combined with a reduction theory to simplify each network description - see \cite{Ott2008, Luke2013, montbrio2015} - and to express the macroscopic phase resetting curve (mPRC) of their oscillatory cycle \cite{Akao2018, kotani2014,Dumont2017}.

The network mPRC is an important causal measure which allows us to use the  weakly coupled oscillator theory \cite{Stiefeljn2015,Ashwin2016} to characterise the inter-network dynamics. 
The fundamental assumption at the core of this theoretical setting is that synaptic projections from one circuit to another must be sufficiently weak. Please note that the weak coupling condition is not on the synaptic connections within each of the circuits, but only across them. 
The weak coupling condition allows one to take advantage of a variety of mathematical techniques and to abbreviate the bidirectionally delayed-coupled spiking circuits description to a single phase equation \cite{Nakao2016, Ermentrout1097}.  This  simplification significantly reduces the complexity of the interacting macroscopic oscillations, making them mathematically tractable, while at the same time capturing crucial principles of phase locking.

As we show below, an analysis of the phase equation sheds lights on the synaptic mechanism enabling circuits with emergent global oscillations to bind together.
We give particular attention to the central role played by the conduction delay in producing symmetry-boken states of activity (with purely symmetric connectivity) , i.e to permit the emergence of a variety of non-symmetric phase lags. Such a collection of phase lags has been suggested to facilitate the control and selection of the information flow through anatomical pathways  \cite{Fries2016}, and conduction delay has been at the core of recent discussion regarding the CTC hypothesis \cite{Fries2015}.
Our final goal is then to show that non-symmetric lags lead to a directed functional coupling between the networks. We indeed show that symmetry-broken states induce with a preferred direction of signal transfer between the networks, and therefore provide theoretical support for the role of oscillations in modulating functional connectivity between cortical circuits \cite{Battaglia2012,Palmigiano2017}.

The paper is structured as follows. First, we present the network and neural model which will be used throughout. We expose the low dimensional system for which we can perform a bifurcation analysis and extract the infinitesimal PRC. From there, we compute the so-called interaction function and reduce the bidirectionally delayed-coupled spiking networks to an unique phase equation. The analysis of the phase equation enables us to make several prediction on the locking states between the emerging oscillations. We support our theoretical findings with extensive numerical illustrations and discuss our results in light of the CTC hypothesis and functional connectivity. Finally, the mathematical techniques are exposed in a detailed Methods section at the end of the paper.

\section*{Results}
\subsection*{The Network and its Reduced Description}

Our generic cortical circuit is assumed to be made up of $N_e$ excitatory cells (E-cells) and $N_i$ inhibitory cells (I-cells) coupled in an all-to-all fashion.  Each cell is described by a well-established conductance-based model - the quadratic integrate-and-fire (QIF), see \cite{Ermentrout:2008} - which is known to capture the essential dynamical features of the neural voltage  \cite{Izi}. The onset of an action potential is taken into account by a discontinuous reset mechanism. Whenever a cut off value $v_{th}$ is reached, the voltage is instantaneously set to $v_r$, a reset parameter. To permit analytical computations, threshold $v_{th}$ and reset $v_r$ are respectively taken at plus and minus infinity \cite{Izi}. The QIF reads 

\begin{equation}
\label{QIF}
\tau \frac{d}{dt}v_j=\eta_j+v_j^2+I ,
\end{equation}%
where $v(t)$ is the neural voltage, $j$ the neuron number, $\tau$ the membrane time constant, $\eta$ the bias current that defines the intrinsic resting potential and firing threshold of the cell  
and finally $I(t)$ the total synaptic current injected at the soma. To account for the network heterogeneity, the intrinsic parameter $\eta$ is distributed randomly according to a Lorentzian distribution:

\begin{equation*}
\mathcal{L}(\eta)= \dfrac{1}{\pi} \dfrac{\Delta}{(\eta - \bar{\eta})^2 +\Delta^2}.
\end{equation*}
Here $\bar \eta$ stands for the mean value taken by the parameter $\eta$ across the population and $\Delta$ is the half-width of the distribution.  Note that the heavy-tailed Lorentzian distribution implies a wide range of intrinsic excitability, i.e. many neurons are not intrinsically oscillating and if they do, they have different firing frequency, as opposed to the classical framework of phasing of coupled oscillators (see \cite{STROGATZ20001} for instance). Indeed, for a null current, the proportion of neurons not being intrinsic oscillator is given by
\begin{equation*}
\int_{-\infty}^{0}   \mathcal{L}(\eta) \, d\eta= \dfrac{1}{\pi} \left( \dfrac{\pi}{2}  - \arctan \left( \dfrac{\bar \eta }{\Delta} \right) \right) ,
\end{equation*}
which can not be zero as soon as there is heterogeneity within the network. Note nonetheless that the proportion will be affected by the synaptic current $I$.

\begin{figure}[t!]
\begin{center}  
   \includegraphics[width= \textwidth]{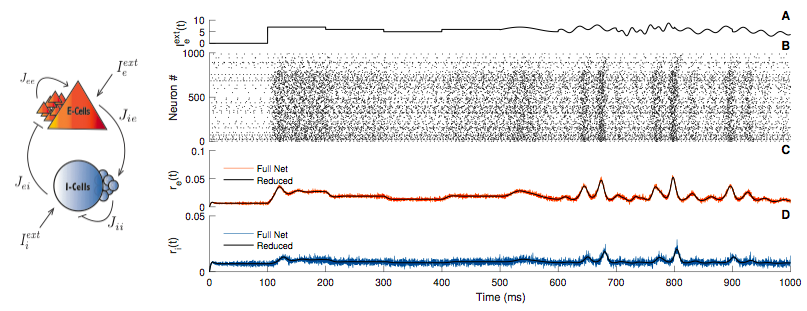}  
   \caption{Comparison between the full network and the reduced system. Left panel: Schematic illustration of a canonical cortical neural network. The parameter $J_{\alpha \beta}$ denotes the connectivity strength of the population $\beta$ onto the population $\alpha$. The external influence on the population $\alpha$ is denoted  $I^{ext}_{\alpha}$. Right panels:  A) Time evolution of the stimulus $I_e^{ext}$ on the E-cells. B) Spiking activity obtained from simulations of the full network, the first $800$ cells are excitatory, the last $200$ are inhibitory. C) Firing rate of the E-cells obtained from simulations of the full network (red line) compared with the reduced system (black line). D) Firing rate of the I-cells obtained from simulations of the full network (blue line) compared with the reduced system (black line). Parameters: $N_e=N_i=5000$; $\Delta_e = \Delta_i=1$; $\tau_e=\tau_i=10$; $\tau_{se}=\tau_{si}=1$; $\bar \eta_e = \bar \eta_i =-5$; $J_{ee}=0$; $J_{ei}=15$; $J_{ii}=10$; $J_{ie}=15$; $I_i^{ext}=0$; $v_{th}=500$; $ v_r=-500$.  }
    \label{fig01}
      \end{center}
\end{figure}

The total synaptic current, $I(t)$ is assumed to be the sum of an external input
$I^{ext}(t)$ that takes into account inputs coming to the cell from sub-cortical structures or nearby cortical networks through lateral connections, and the synaptic inputs $s_e$ and $s_i$ which models the effect of recurrent connexions within the circuit:

\begin{equation*}
I= I^{ext}+  \tau s_e - \tau s_i.  
\end{equation*}%
The synaptic current, $s(t)$, depends on the synapse type, for the excitatory synapse,  we have

\begin{equation*}
\tau_s \frac{d}{dt} s_e=-s_e +J_{e} r_e ,  
\end{equation*}%
respectively for the inhibitory synapse,
\begin{equation*}
\tau_s \frac{d}{dt} s_i=-s_i +J_{i} r_i . 
\end{equation*}%
Here, $\tau_s$ the synaptic time constant, $J$ the synaptic strength - see Fig. \ref{fig01} - and $r(t)$ the population firing rate. For the E-cells, we have:

\begin{equation*}
r_e(t)= \dfrac{1}{N_e}  \sum_{k=1}^{N_e}\sum_{f}\delta(t-t_f^k) ,
\end{equation*}
and for the I-cells, we have:

\begin{equation*}
 r_i(t)= \dfrac{1}{N_i}  \sum_{k=1}^{N_i}\sum_{f}\delta(t-t_f^k),
\end{equation*}
where $\delta$ is the Dirac mass measure and $t_f^k$ are the firing time of the neuron numbered $k$.

To get a clear picture of how the synaptic structure shapes the firing patterns, we take advantage of a thermodynamic approach combined with a reduction method. The thermodynamic framework produces a single average system written in term of partial differential equations that is valid in the limit of an infinitely large number of neurons \cite{Deco2008}. The reduction method allows further simplification and breaks down the mean-field system into a small set of differential equations \cite{montbrio2015, Luke2013}. In our case, see Method for more details about the derivation, the low dimensional dynamical system reads:

\begin{equation}\label{RedE}
\left\lbrace
\begin{split}
&\tau_e  \frac{d}{dt}r_e = \dfrac{\Delta_e}{\pi \tau_e} +2  r_e V_e \\ 
&\tau_e  \frac{d}{dt}V_e = V_e^2 +\bar \eta_e  +I_e  - \tau_e^2 \pi ^2 r_e ^2,\\
&\tau_s \frac{d}{dt} s_{ee}=-s_{ee} +J_{ee} r_e ,  \\
&\tau_s \frac{d}{dt} s_{ei}=-s_{ei} +J_{ei} r_i ,  
 \end{split}\right.
\end{equation}
and for the I-cells:
 \begin{equation}\label{RedI}
\left\lbrace
\begin{split}
&\tau_i  \frac{d}{dt}r_i = \dfrac{\Delta_i}{\pi \tau_i} +2  r_i V_i \\ 
&\tau_i  \frac{d}{dt}V_i = V_i^2 +\bar \eta_i  +I_i - \tau_i^2\pi ^2 r_i ^2.\\
&\tau_s \frac{d}{dt} s_{ie}=-s_{ie} +J_{ie} r_e ,  \\
&\tau_s \frac{d}{dt} s_{ii}=-s_{ii} +J_{ii} r_i ,  \\
\end{split}\right.
\end{equation}
In here, $V(t)$ represents the mean voltage of the population, while $r(t)$ still stands for the firing activity. Note that the two systems are coupled via the expression of the total current arriving on each sub-population:

\begin{equation*}
I_e= I_e^{ext}+  \tau_e s_{ee} - \tau_e s_{ei}, 
\end{equation*}%
and
\begin{equation*}
I_i= I_i^{ext}+  \tau_i s_{ie} - \tau_i s_{ii}.  
\end{equation*}%

The numerical simulation presented in Fig. \ref{fig01} compares the dynamics of the full network with the low dimensional system (\ref{RedE})-(\ref{RedI}). It shows the time evolution of the external stimulus in the first panel (Fig. \ref{fig01}A), whereas the second panel gives the spiking activity obtained from a simulation of the full network (Fig. \ref{fig01}B). In the subsequent panels (Fig. \ref{fig01}C-D), the firing rate given by the reduced description is compared with the firing rate obtained from network simulations.

The perfect agreement between the population activities convinced us that the reduced dynamical system captures the fundamental aspects of the population firing rate.  Of course, such a reduced description provides an efficient way to carry out a study of the circuit since it can be simulated very quickly and it is amenable to mathematical analysis.

\subsection*{Emerging Rhythms and Phase-Resetting Curve}

 \begin{figure}[t!]
\begin{center}  
   \includegraphics[width= \textwidth]{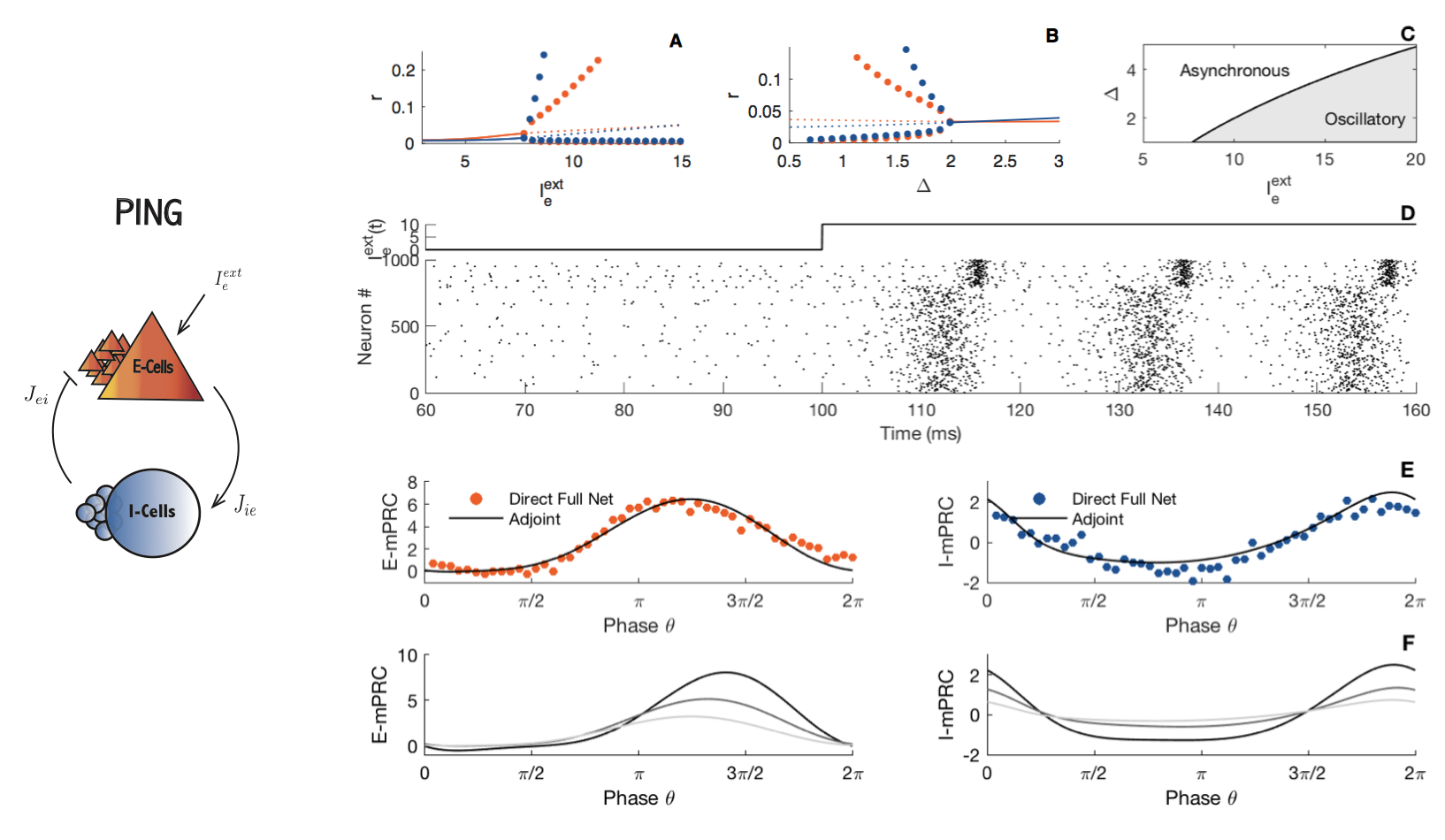}  
   \caption{The PING interaction. Left panel: Schematic illustration of the PING (Pyramidal Interneuron Network Gamma) interaction. The parameters have similar properties than Fig. \ref{fig01}. Right panels: Nonlinear analysis of the The PING interaction. A-B) Bifurcation diagrams. The blue line, (respectively the red line), corresponds to the steady state of the inhibitory cells, (respectively the excitatory cells) while dots correspond to limit cycles.  C) Stability region. D) The stimulus and corresponding raster plot of the spiking activity. E) Comparison between simulated and calculated mPRCs. The black line illustrates the analytical adjoint method while dots indicates direct perturbations of the full network. Red dots, perturbations are made on the E-cells, second row, with the blue dots, perturbations are made on the I-cells. F) PRC shape as a function of parameters obtained via the adjoint method for different values of the external current $I_e^{ext}=9,10,11$. Parameters are as in Fig. \ref{fig01}, except  $J_{ee}=0$; $J_{ei}=15$; $J_{ii}=0$; $J_{ie}=15$ and for the panel B)  $I_{ext}^e=10$. Direct perturbations in panels E) are made with a square wave current pulse (amplitude $10$, duration $0.5$). } 
    \label{fig04}
      \end{center}
\end{figure}

To understand how the emergent network gamma oscillations can phase lock, it is essential to first consider their basic 
underlying mechanisms. To gain insights, a nonlinear analysis of the reduced system is performed. This enables us to reveal how the inhibitory feedback loop renders possible the emergence of macroscopic rhythms. Two processes can be described: PING and ING \cite{Bartos2007}.

In the PING (Pyramidal Interneuron Network Gamma) interaction, see Fig. \ref{fig04}, the underlying synaptic machinery involves an interplay between the pyramidal cells and the fast-spiking cells.
For a chosen set of connectivity parameters, the dynamical system exhibits a Hopf bifurcation (Fig. \ref{fig04}A), such that, enhancing the external stimulus upon the pyramidal cells induces a graded progression toward an oscillatory regime. Note that this rhythmic regime disappears as the network heterogeneity is expanded (see Fig. \ref{fig04}B-C).  
The rhythmic transition is illustrated with a simulation displayed in Fig. \ref{fig04}D. A self-sustained oscillatory regime emerges as soon as the E-drive is strong enough.
Of course, the presence of a Hopf bifurcation in the system should be put in relation with the seminal work of Wilson and Cowan \cite{Wilson1972}.

\begin{figure}[t!]
\begin{center}  
   \includegraphics[width=  \textwidth]{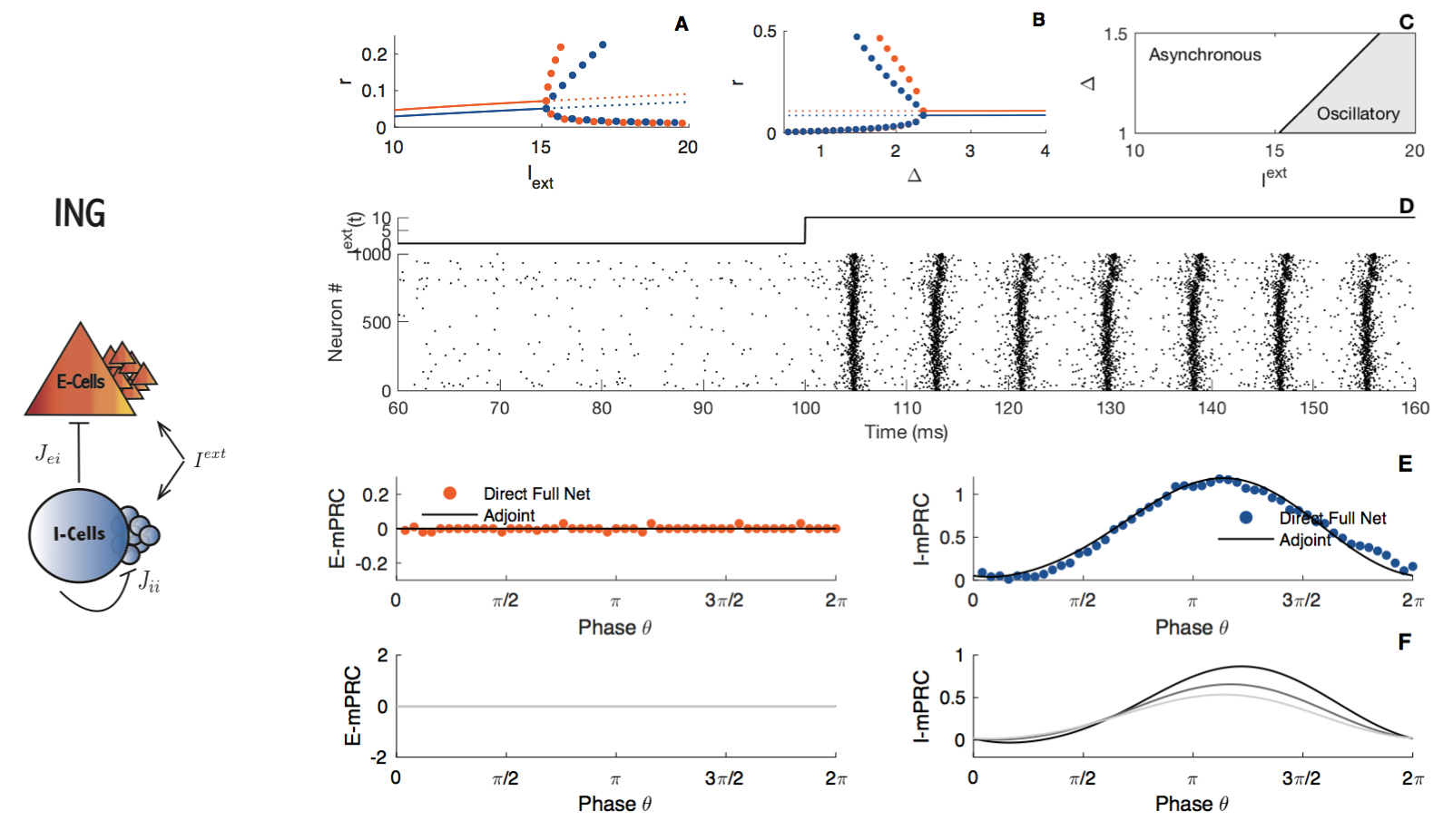}  
   \caption{The ING interaction. Left panel: Schematic illustration of the ING (Interneuron Network Gamma) interaction. The parameters have similar properties than Fig. \ref{fig01}. Right panels: Nonlinear analysis of the The ING interaction. A-B) Bifurcation diagrams. The blue line, (respectively the red line), corresponds to the steady state of the inhibitory cells, (respectively the excitatory cells) while dots correspond to limit cycles.  C) Stability region. D) The stimulus and corresponding raster plot of the spiking activity.  E) Comparison between simulated and calculated mPRCs. The black line illustrates the analytical adjoint method while dots indicates direct perturbations of the full network. Red dots, perturbations are made on the E-cells, second row, with the blue dots, perturbations are made on the I-cells. F) PRC shape as a function of parameters obtained via the adjoint method for different values of the external current $I^{ext}=24, 25, 26$. Parameters are as in Fig. \ref{fig01}, except  $J_{ee}=0$; $J_{ei}=10$; $J_{ii}=15$; $J_{ie}=0$ and for the panel B)  $I_{ext}=25$. Direct perturbations in panels E) are made with a square wave current pulse (amplitude $10$, duration $0.5$).}
    \label{fig05}
      \end{center}
\end{figure}

In the ING (Interneuron Network Gamma) interaction, see Fig. \ref{fig05}, the mechanism requires an inhibitory feedback from fast-spiking cells onto themselves and the rhythm arises from this interconnected inhibitory network which in turn defines the excitatory spike times. The nonlinear analysis reveals a Hopf bifurcation as the external drive is raised (see  Fig. \ref{fig05}A). Again, this rhythmic regime disappears with too much heterogeneity (see Fig. \ref{fig05}B-C). The network activity undergoes a transition from an asynchronous regime toward an oscillatory which is displayed in Fig. \ref{fig05}D.  Interestingly, the ING behavior can not emerge within the traditional rate equation proposed by Wilson and Cowan \cite{Wilson1972}, see \cite{Devalle2017} for a more complete discussion. 

Note finally the frequency difference between the PING and the ING rhythm. The two interaction models are then seen as canonical descriptions of the low and fast gamma oscillations, PING for low gamma range and ING for fast gamma spectrum.

Over the past decades, the Phase Resetting Curve (PRC) has become one of the fundamental concepts in theoretical neuroscience. Its usefulness has been reviewed in multiple 
papers  \cite{Smeal2407,Stiefeljn2015,Ashwin2016,Nakao2016} and its outcomes are expected to impact our understanding of brain rhythms \cite{Canavier2015}. 
PRC measures the effects of transient stimuli upon oscillatory systems and can be obtained experimentally \cite{reyes1993,Akam2012,Stiefel2009,Phoka2010}.

In our case, the application of a short depolarizing current to the network affects the spiking activity, and the macroscopic oscillation shifts in time.  The induced phase shift depends on the perturbation strength but also on the phase at which the perturbation is presented. It can either delayed or advanced  depending on the onset phase of the perturbation

The PRC results in plotting the advance or delay with respect to the phase onset at which the perturbation is made. Doing so, it quantifies the effect of the perturbation on the macroscopic oscillation. For the cortical network under consideration, several PRCs coexist at the same time depending on where the depolarizing input is applied.

In the limit of short, weak perturbations, the shift in timing can be described by the so-called infinitesimally PRC (iPRC). 
The iPRC is mathematically expressed by a linear differential system, known as the the adjoint system \cite{Brown2004}. This method can be applied to the low dimensional system (\ref{RedE})-(\ref{RedI}) and a semi-analytical expression of the iPRC be obtained. Assuming that the reduced E-I system (\ref{RedE})-(\ref{RedI}) has a stable limit cycle, 
we find that, see Method for more details, the iPRC $Z(t)$ is a periodic vector that is a solution of the adjoint equation

\begin{equation}\label{iPRC}
-\frac{d}{dt}Z(t) = \mathcal{M}(t)^T \cdot Z(t),
\end{equation}
where the matrix $\mathcal{M}(t) $ is given by a linearization of the E-I system (\ref{RedE})-(\ref{RedI}) around the limit cycle, see Method for its precise expression. 

When perturbations made to the network are sufficiently small, the PRC becomes proportional to the iPRC  \cite{kotani2014,Kotani2012,Nakao2014}. 
We present in Figs. \ref{fig04}E and \ref{fig05}E and  the iPRC obtained via a simulation of the adjoint system (\ref{iPRC}) compared with direct perturbations made on the spiking network.  The blue line, (respectively the red line), corresponds to the iPRC of the excitatory synapse of the I-cells (respectively the E-cells).

From the simulations and semi-analytical expression of the PRC we can classify the PING and ING rhythms as having different PRC types, i.e. as having different rhythmic properties. For the PING dynamics, see Fig. \ref{fig04}E, a biphasic shape of the PRC is observable when perturbations are made on the I-cells. In contrast, when perturbations are on the E-cells, the PRC is monophasic. This is a classification already observed in a previous work where the synaptic dynamic was neglected and considered to be instantaneous \cite{Dumont2017}. 

Regarding the ING pattern, see Fig. \ref{fig05}E, the PRC is monophasic for perturbation targeting the I-cells. The PRC is null when perturbations are made onto the pyramidal cells, which means that any perturbations will die out after a few cycle. This comes without a surprise since in the ING interaction, pyramidal cells do not play a part in the emergence of the oscillations.

PRCs are thus quite different between the ING  and the PING oscillations. This is because the contribution of the cell type to the rhythmic behavior is largely different in the ING and PING mechanisms.
The PRC difference between the ING and the PING oscillations have been investigated in very recent work \cite{Akao2018}. From there, we can explore the consequences of differences of locking regimes to periodic pulsatile stimuli, and their result supports that the origin of the cell-type-specific response, already experimentally observed \cite{Cardin2009}, comes from the different entrainment properties \cite{Akao2018}.

Similarly, we can also question the network sensitivity to perturbation to the excitatory cells or to the inhibitory cells from the difference in amplitudes of the respective macroscopic PRCs. As we shall see in Fig. \ref{fig04}F and in Fig. \ref{fig05}F, the PRC amplitude strongly depends on parameters such as the external current. This can be intuitively interpreted as a dependence of stability of the macroscopic oscillation.

\subsection*{Phase equation}

\begin{figure}[]
\begin{center}  
\includegraphics[width=\textwidth]{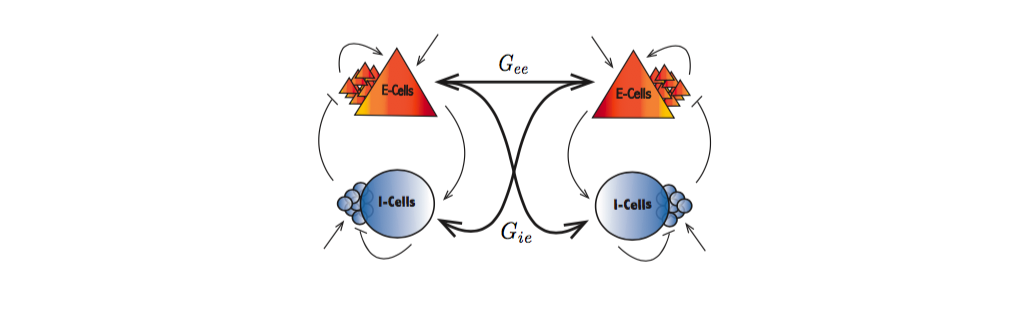} 
   \caption{The bidirectionally neural circuits. Top panel: The two coupled circuits. The parameter $G_{\alpha \beta}$ denotes the connectivity strength of the population $\beta$ of one network onto the population $\alpha$ of the other circuit. The intrinsic parameters are unchanged and similar within each network as presented in Fig. \ref{fig01}.  }
    \label{fig08}
      \end{center}
\end{figure}

We now turn our investigation to the dynamical emergence of phase synchrony across multiple networks (as reflecting multiple brain regions). While we do not aim at studying any specific brain interaction, however, the structure that is shown in Fig. \ref{fig08} reflects the architecture of many communicating cortical and sub-cortical areas where information transmission is at play \cite{Deco2010,Barardi2014}. Each network is assumed to be made up of interacting pyramidal cells and interneurons as presented in the previous sections (see Fig. \ref{fig01}). Since interneurons are known to wire on local scale, the synaptic projection from one circuit to another is made via the pyramidal cells only. A delay is added to account for finite transmission speeds and synaptic time-courses across circuits. Importantly we note that the considered structural motif is symmetric, i.e. it is stable under permutation of the two cortical networks.

Note that locking across gamma oscillations appears within the same frequency range, we will thus focus our study on two interacting schemes: the PING-PING interaction and the ING-ING interaction. The two mechanistic models of gamma generation having different oscillatory regimes, the interaction PING-ING would lead to a cross-frequency coupling. First, it is far beyond the scope of this paper to investigate the coherence between slow and fast oscillations, second, we note that, under our knowledge, cross-coupling among slow and fast gamma  has not been observed so far.

Our whole analysis of phase locked states is based on the assumption that synaptic interactions across circuits are sufficiently weak. Such an assumption, which guarantees that the perturbed macroscopic oscillation remains close to the unperturbed one, allows to place our study within the framework of weakly coupled oscillators  \cite{Nakao2016, Ermentrout1097}. We emphasize that within each circuit, neurons are not weakly coupled. The assumption of weak coupling is only made upon the projection from one circuits to another. Within this framework, see Methods, the bidirectionally delayed-coupled neural circuits reduce to a single phase equation:

\begin{equation*}
\dfrac{d}{dt} \theta(t) = G(\theta(t)),
\end{equation*}
where $\theta(t)$ is the phase difference (or phase lag) between the circuits and the $G$-function is the odd part of the shifted interaction function (the so-called $H$-function) expressed via the PRC (see Method):

\begin{equation*}
\begin{split}
H(\theta) = & \dfrac{G_{ee}}{T}\int_0^T Z_{s_{ee}}(s)r_e(s-\theta) \, ds  \\
& + \dfrac{G_{ie}}{T}\int_0^T Z_{s_{ie}}(s)r_e(s-\theta) \, ds ,
\end{split}
\end{equation*} 
where $T$ is the oscillation period and $G_{\alpha \beta}$ denotes the connectivity strength of the population $\beta$ of one network onto the population $\alpha$ of the other circuit, see Fig  \ref{fig08}. Note the involvement of the synaptic component of the PRC $Z_s(t)$ and the firing rate of the E-cells $r_e(t)$ all along the oscillatory cycle.

The $G$-function is essential for our study since it conveys knowledge about the phase-locking mode between the coupled circuits. Indeed, the zeros of the $G$-function correspond to steady states phase lags. The stability of a locking mode is conditioned by a negative slope ($G'(\theta) < 0$), while a positive slope ($G'(\theta) > 0$) implies instability.

\subsection*{Locking modes}

\begin{figure}[t!]
\begin{center}  
      \includegraphics[width= \textwidth]{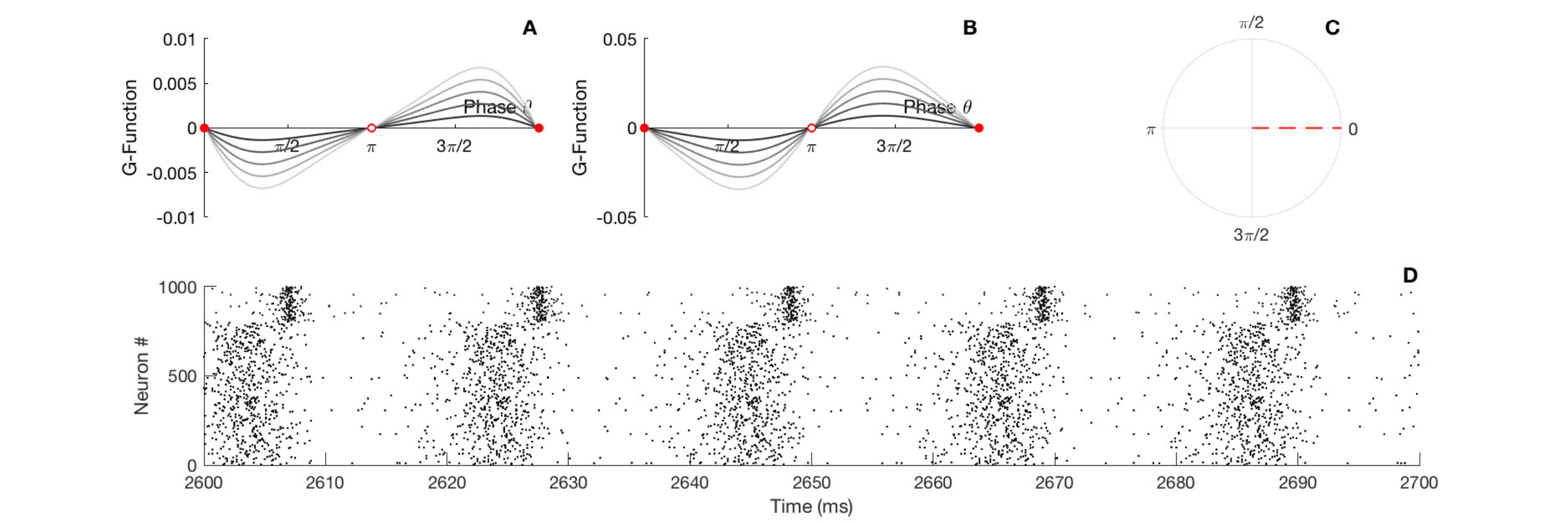}  
      \includegraphics[width= \textwidth]{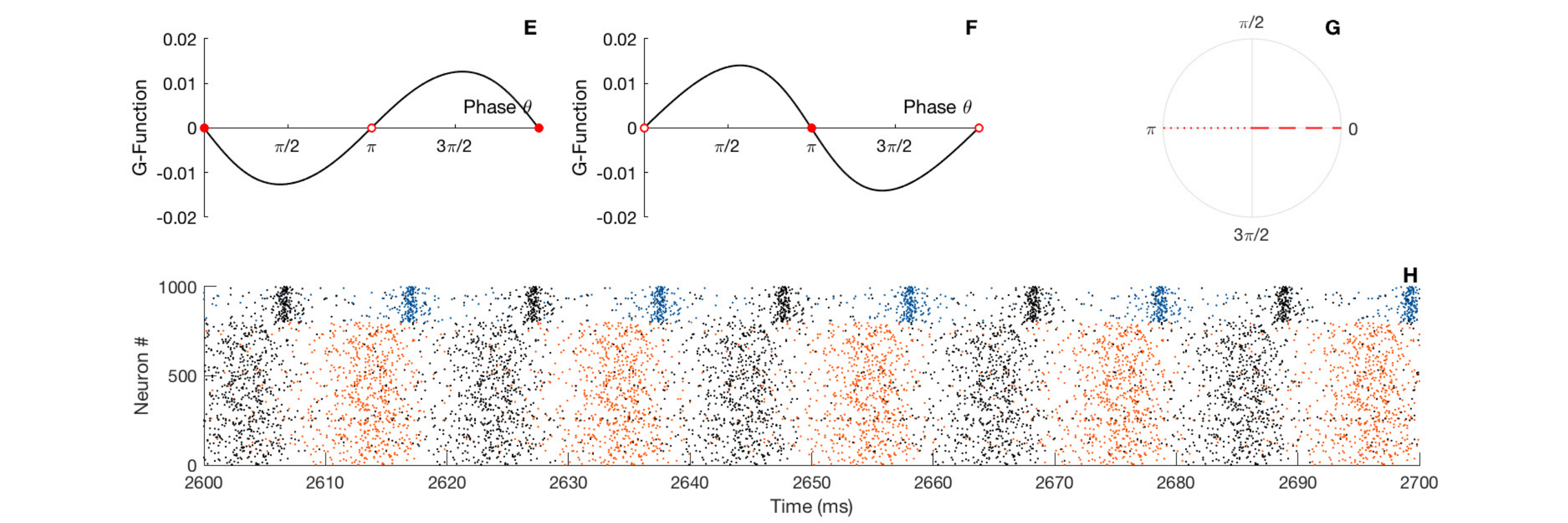}  
   \caption{  locking modes of PING. A) The panel gives the $G$-function for different parameter $G_{ee}$ when $G_{ie}=d=0$.  B) The panel gives the $G$-function for different parameter $G_{ie}$ when $G_{ee}=d=0$. The circles are filled for stable fixed point and empty for the unstable points.  C) Resulting locking mode when there is no delay.  D) Raster plot of the spiking activity of the two neural networks, black dots indicate the spike timing of the first network, colored dots indicate the spike timing of the second network. 
  E) The panel gives the $G$-function for $d=2$, $G_{ee}=0.1$;  $G_{ie}=0.5$;.  F) The panel gives the $G$-function for $d=10$; $G_{ee}=0.1$;  $G_{ie}=0.5$;. The circles are filled for stable fixed point and empty for the unstable points.  G) Resulting locking mode for short and large delay.  H) Raster plot of the spiking activity of the two neural networks, black dots indicate the spike timing of the first network, colored dots indicate the spike timing of the second network. Parameters are as in Fig. \ref{fig06}. }
    \label{fig09}
      \end{center}
\end{figure}

To disentangle the synaptic mechanisms responsible for the dynamical emergence of cross-network phase locking, we first fix the delay to zero and focus our study on the effect of coupling weights. To put it in mathematical terms, we investigate the location of the zeros of the $G$-function with respect to the coupling strengths when the parameter $d$ is set to zero. 
As we see from  Fig. \ref{fig09}, modification in the network parameters affects the shape of the $G$-function. This is made for the PING interaction. The zeros of the $G$-function are located at the in-phase (synchrony) and anti-phase locking (anti-synchrony) mode. The anti-phase state is nonetheless unstable.

We therefore expect the in-phase synchrony mode to emerge from the dynamic of the bidirectionally coupled circuits. This is the case for a cross-coupling targeting exclusively the E-cells ($G_{ie}=0$, Fig. \ref{fig09}A) or the I-cells only ($G_{ee}=0$, Fig. \ref{fig09}B). Since in the general case, the interaction function will result in a linear superposition of the two previously mentioned possibilities, in the non-delayed coupling scenario, only a perfect zero lag synchrony can be expected, see Fig. \ref{fig09}C. 
We illustrate this prediction by showing the network rasters in Fig  \ref{fig09}D. The black dots correspond to the first network, whereas the colored dots to the second circuit. The spiking activity of the two circuits oscillate in phase, i.e. the two raster plots are synchronized at zero lag and thus perfectly overlap and that is the reason why only black dots can be seen. Simulation and theoretical prediction are in perfect agreement. As we can see, despite its vast simplifications, the phase equation yields amazingly accurate predictions.

The fact that two oscillatory networks (two oscillators) synchronize at zero lag when delay is neglected was to be expected. 
However, in real settings, neuronal signals travel at finite speeds across the brain and a wide range of delays between neuronal populations has been reported  \cite{delay}. How the presence of transmission delay reshapes the phase relationship between macroscopic oscillations has remained elusive so far. This is a central issue since recent studies have proposed an updated formulation of the CTC hypothesis where delay between communicating sites plays a critical role \cite{Fries2015}.

To put it into a mathematical perspective, we expect that distinct delays lead to different fixed-points in the $G$-function, and to illustrate this expectation, we plot the $G$-function obtained for two different delays (Fig. \ref{fig09}E-F). As we can see, the stability of the locking modes are reversed, and the anti-phase mode, which was unstable, becomes  stable. In contrast, the in-phase mode turns into an unstable state. Two possible phase locking modes are then possible, the in phase mode for short delay and the anti-phase mode for large delay (Fig. \ref{fig09}E-F). We illustrate this prediction by showing the network rasters in Fig  \ref{fig09}H. As we can see, for large delay value, the spiking activity of the two circuits oscillate in an out of phase mode.

\begin{figure}[t!]
\begin{center}  
   \includegraphics[width= \textwidth]{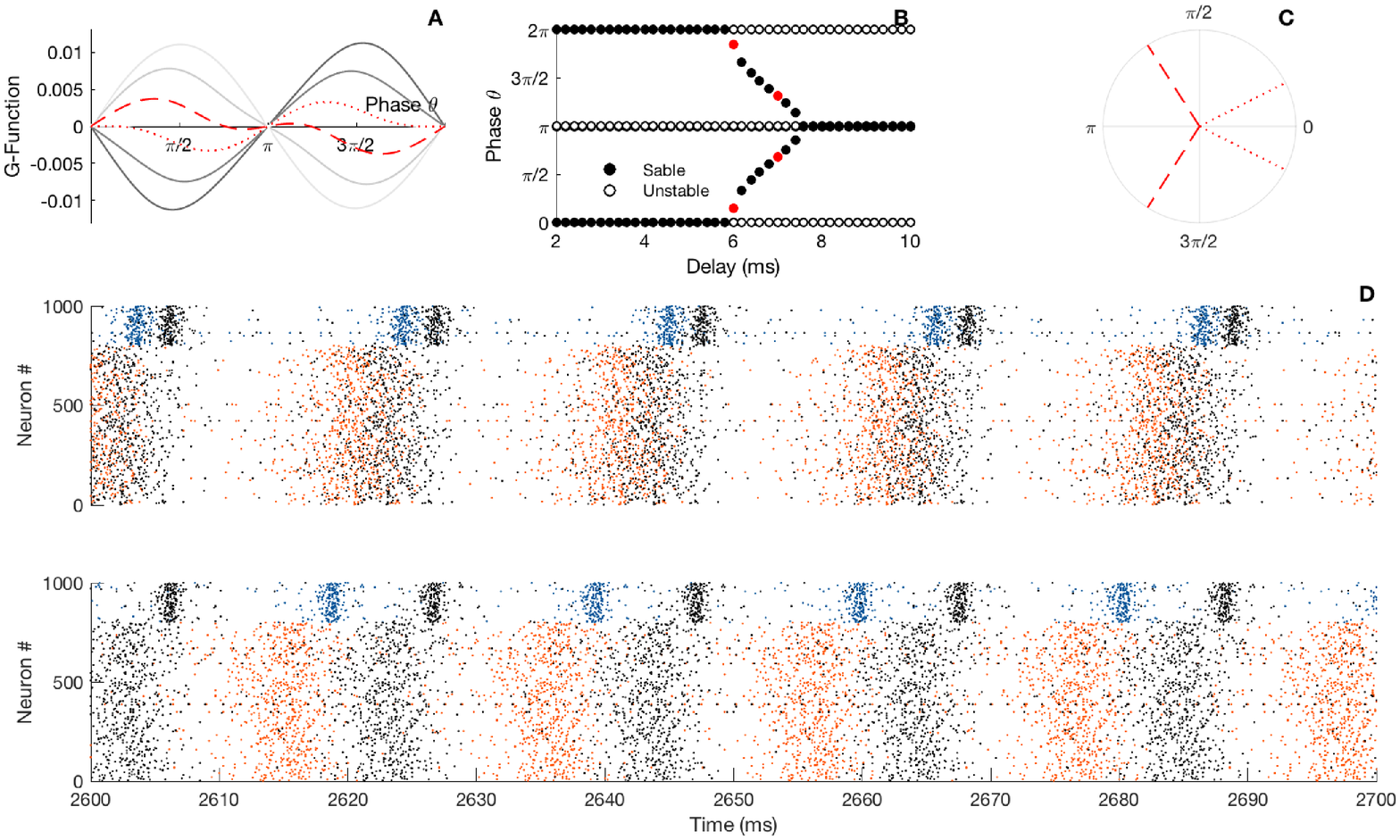}  
      \includegraphics[width= \textwidth]{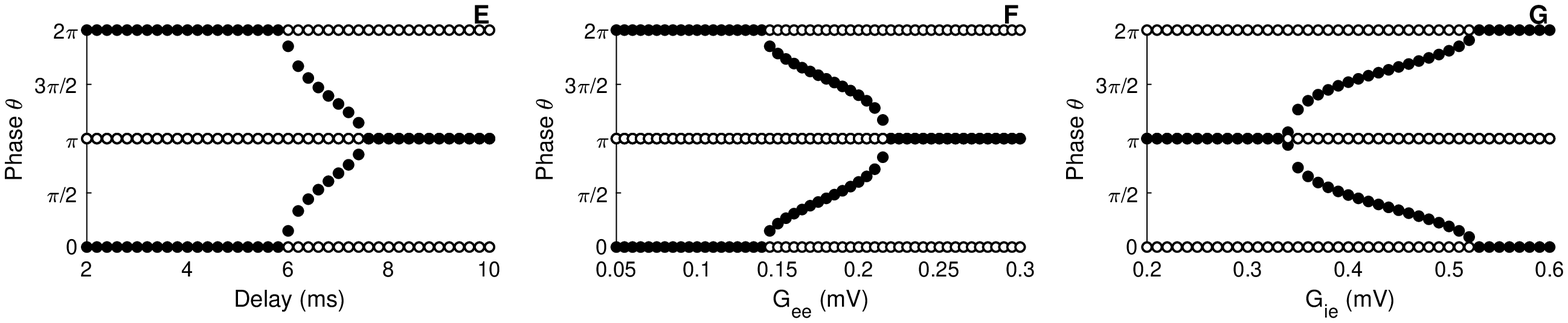} 
   \caption{Bifurcation analysis of locking modes with respect to delay for the PING. A) The panel gives the $G$-function for different parameter $d$.  B) Bifurcation analysis. The circles are filled for stable fixed point and empty for the unstable points, red dots represent the illustrations parameters C) Resulting locking modes.  D) Raster plot of the spiking activity of the two neural networks, black dots indicate the spike timing of the first network, colored dots indicate the spike timing of the second network. Parameters are as in Fig. \ref{fig06} with $G_{ee}=0.1$;  $G_{ie}=0.5$;  A) $d=6$, B) $d=7$.  }
    \label{fig11}
      \end{center}
\end{figure}

We push further the analysis by investigating the transition between the two phase locking modes.
In Fig. \ref{fig11}A  we plot the $G$-function obtained several delays, black lines correspond to small delays while grey lines to bigger ones. A continuous deformation is seen, leading the $G$-function to slip over the phase-axis. 
To get a better visualization, we plot a bifurcation diagram (Fig. \ref{fig11}B) which gives access to the phase mode with respect to parameter change. Such a diagram is very helpful to anticipate the locking states in the bidirectionally delayed-coupled networks. We note that the stability of the in-phase mode is kept for small delays. On the other hand, for larger transmission time, a switch of stability between the in-phase and anti-phase locking modes is observed. Importantly, a wide region of delays for which the phase lag goes over all the possibilities appears in the diagram. This result confirms the role of delay in the emergence of a complete variety of phase shifts across gamma interaction in the cortex \cite{Canolty2010,Fries2016}.

In Fig. \ref{fig11}D we validate this theoretical prediction by showing rasters of the spiking circuits that reflects the modulation of the emerging phase lag by the delay.  As we see from Fig  \ref{fig11}D, the spiking activity of the two networks oscillate with a small phase lag. Increasing slightly the delay leads the spiking activity of the two networks to oscillate with a bigger phase lag. Simulation and theoretical prediction are again in perfect agreement. This result shows that it is normal to observe persistent phase relationship across time that are so diverse across space. 

As already noticed in \cite{Battaglia2007}, this case corresponds to a spontaneous symmetry breaking. We talk about symmetry breaking because those variety of phase lag states do not share the symmetric feature with the
full system.  Note that when the delay is kept fixed, and sufficiently large, a variation of the synaptic strength onto the E-cells in Fig. \ref{fig11}F leads to a transition from the in-phase state to the out-of-phase locking. As a part of this transition, a variety of stable phase lags appear.  A reverse situation is depicted in Fig. \ref{fig11}G: when the coupling onto the I-cells is varied the in-phase mode transitions to an anti-phase mode. As we can  see from these diagrams, we can  tune the phase shifts across brain oscillations at least for the PING rhythm. Indeed, this only possible for the PING rhythm since in that case two non-zero PRCs exist at the same time.

\begin{figure}[t!]
\begin{center}  
   \includegraphics[width= \textwidth]{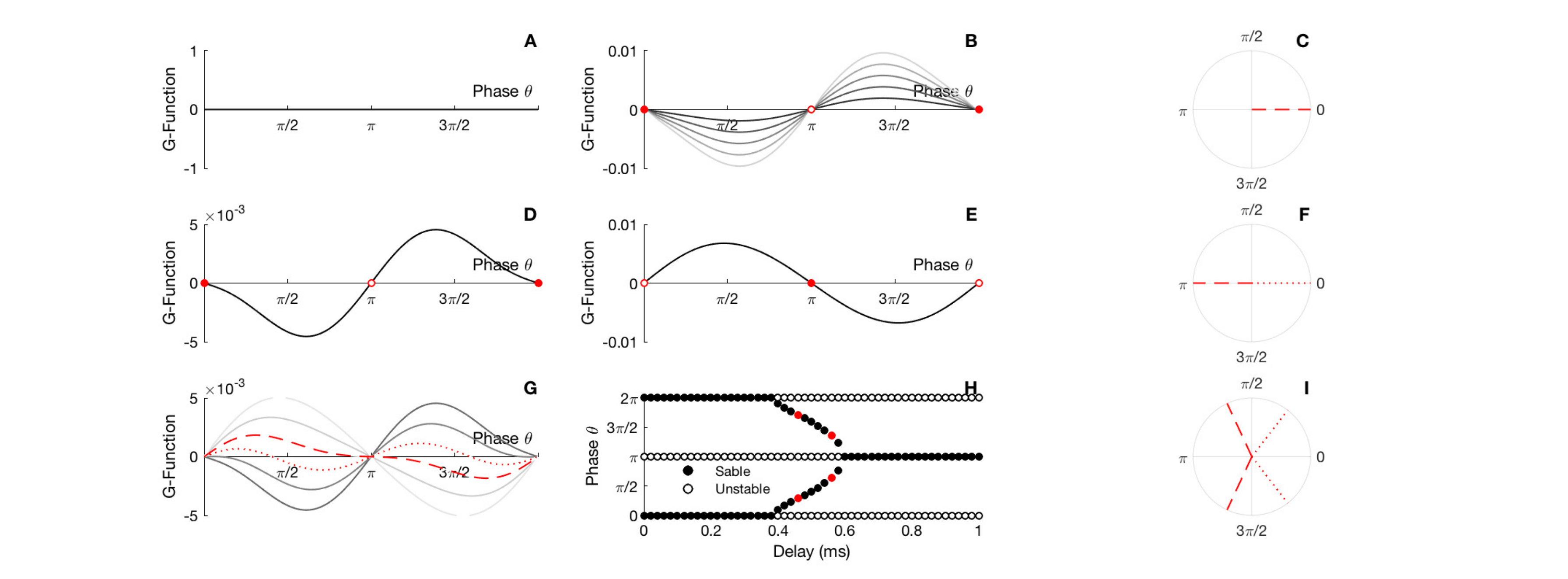}  
   \caption{Locking modes of the ING. Effect of coupling strength on locking modes without delay. A) The panel gives the $G$-function for different parameter $G_{ee}$ when $G_{ie}=d=0$.  B) The panel gives the $G$-function for different parameter $G_{ie}$ when $G_{ee}=d=0$. The circles are filled for stable fixed point and empty for the unstable points.  C) Resulting locking mode when there is no delay.
   D) The panel gives the $G$-function for $d=0.1$.  E) The panel gives the $G$-function for $d=0.6$. The circles are filled for stable fixed point and empty for the unstable points.  F) Resulting locking mode for short and large delay.     G) The panel gives the $G$-function for different parameter $d$.  H) Bifurcation analysis. The circles are filled for stable fixed point and empty for the unstable points, red dots represent the illustrations parameters I) Resulting locking modes. Parameters are as in Fig. \ref{fig05} with $G_{ee}=0.$;  $G_{ie}=0.3$;  A) $d=0.2$, B) $d=1$.  }
    \label{fig12}
      \end{center}
\end{figure}

A similar situation emerges for the ING interaction. In Fig. \ref{fig12}, we show the interaction function and corresponding locking modes. While short delays induces only an in-phase locking mode Fig. \ref{fig12}C-F, larger delays will reverse the interaction function and induce an out of phase locking scheme Fig. \ref{fig12}E-F. Once again, notice the spontaneous symmetry breaking implying the existence of a variety of phase lags for moderate values of the delay Fig. \ref{fig12}G-H-I. Not that for the ING-ING interaction, modification of the synaptic coupling $G_{\alpha \beta}$ will not affect the locking modes since the coupling is through the pyramidal neurone and these do not affect the macroscopic oscillatory phase.


\subsection*{Emerging causal directionality}

We now turn to the functional role that could be supported by a breaking of symmetry between the two structurally symmetrically connected brain circuits, each with a emergent oscillatory activity. Recent studies have associated spontaneous symmetry breaking with an unidirectional effective transfert of information despite a symmetric structural connectivity motif \cite{Battaglia2012,Palmigiano2017}. We therefore expected to characterize a similar phenomenon but with a direct causal measure . 
To prove that there is indeed a causal directionality of signaling, we compute a global PRC of the full delay system. Our intention is to measure how an impact on one of the two networks affects the other circuit and the system as a whole. In the symmetry broken state, there is a leader circuit and a follower circuit. We posited that should we find that the global PRCs are identical for perturbations to either the leader or the follower, signal transfer is symmetric, should the two PRCs differ, signal transfer is directed. While we seek a fully analytical approach, the difficulty with the presence of the delay is that it hard to offer an analytical treatment.  An analytical method for such a global PRC is not easy to implement and its convergence is not guarantee. We therefore follow a semi-analytical approach, where we use the direct method to compute the global PRC for the reduced model, which makes the computations efficient (see Method Eqs. (\ref{RedE1})-(\ref{RedI2})).  We this perturb one network (the leader or the follower) and observe the resulting asymptotical phase shift of the second network.

\begin{figure}[t!]
\begin{center}  
      \includegraphics[width=\textwidth]{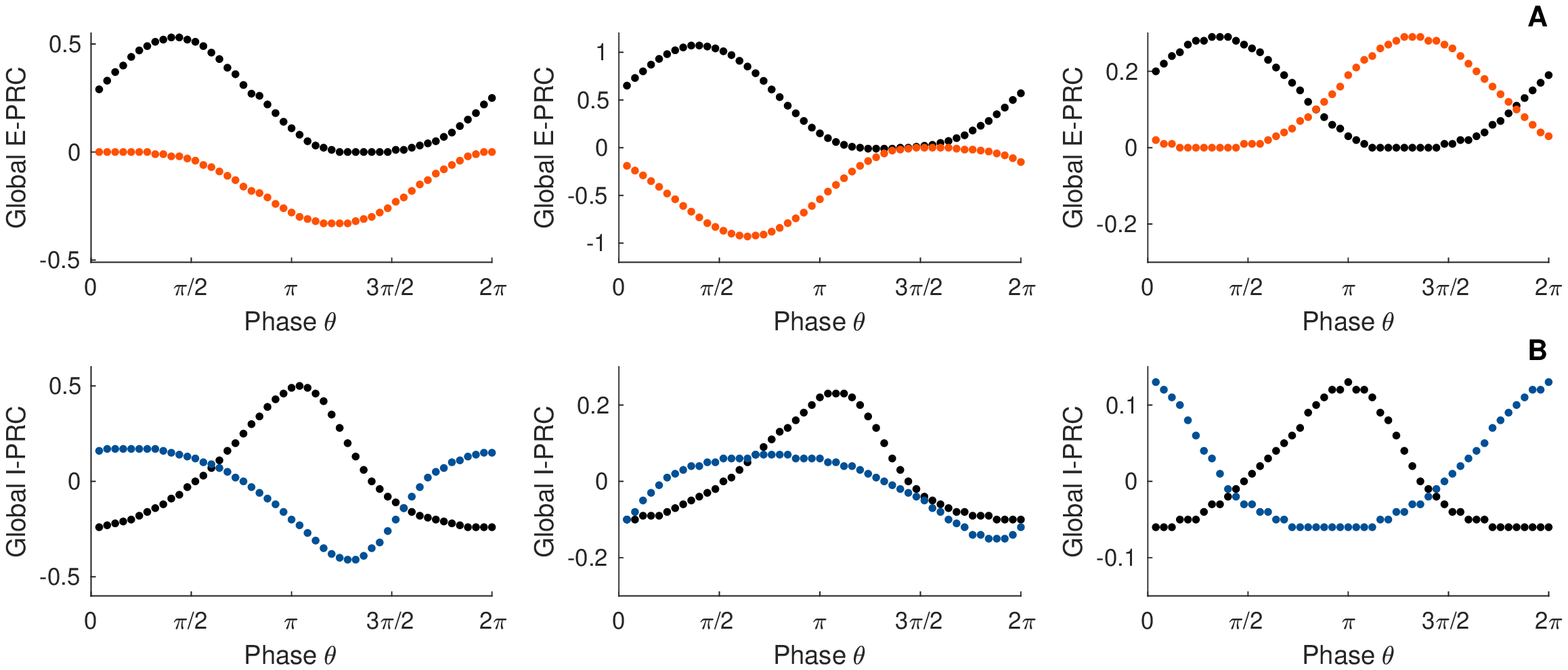}  
   \caption{Global PRC and emerging directionality. Black dots illustrate direct perturbations of the leader, color dots indicate direct perturbations of the follower. A) Perturbations are made on the E-cells. B) Perturbations are made on the I-cells. The network parameters are the same as in Fig. \ref{fig04} with $I_e^{ext}=10$, direct perturbations are made with a square wave current pulse (amplitude $1$, duration $0.2$), the first panels $d=6$, the second panel $d=7$, the third panel $d=10$.}
    \label{fig14}
      \end{center}
\end{figure}

Fig. \ref{fig14} illustrates the perturbation effects. As expected, when the two networks are in phase, perturbing one or the other has similar outcomes. When the two networks are out of phase, the resulting global PRCs are only shifted with respect to one another. This is a natural consequence of the symmetry in the oscillatory modes of the macroscopic oscillations. The most interesting scenario is when the resulting phase locking mode is not symmetric.  In this situation, perturbing the leader or the follower does not give the same phase shift. As we can see, PRCs are almost reverse, i.e. while a perturbation of the leader induces a phase advance, a perturbation on the follower implies a phase delay. Importantly, the amplitude of the PRCs have also different order of magnitude. Perturbation of the leader have stronger impact than on the follower. Furthermore, we see that for each of the perturbations, phase shifts depend on the phase at which the external "signal" arrives: e.g. in cases where an excitatory perturbation on either networks advances the oscillations, and phases where perturbing the leader advances the phase, while exciting the follower delays the oscillation. As a note, it has been previously shown that the post-stimulus spike-time histogram (PRC) can be directly related to the PRC \cite{Gutkin2005,Ermentrout2007}. Hence, the asymmetric PRCs for the leader and the follower predict non-symmetric PSTHs, once again giving a direct and causal measure of how broken-symmetry states can induce a directional functional connectivity despite complete structural symmetry.

 In summary, we can interpret our results giving a causal directionality in the communication between the two circuits which has been recently showed using  using correlative statistical measure such as transfer entropy  \cite{Battaglia2012,Palmigiano2017}.

\section*{Discussion}

The omnipresence of oscillations in the brain gives significant support to the hypothesis that rhythmic firing patterns are well suited to specific cognitive functions \cite{Buzsaki2004,buzsaki2006rhythms}.
In particular, recent physiological experiments have indeed acknowledged that coherent gamma rhythms play a determinant part in the transfer of information across cortical areas \cite{Fries2005,fries2009neuronal,Fries2015}. 
As this communication depends on stable phase-relationships between the oscillatory cortical networks, a key question has been to determine the conditions under which two oscillatory brain circuits phase lock, what is the resulting phase lag between them and how the phase lag relates with delay and synaptic couplings \cite{kopell2014}.

Here, we have outlined and tested a new analytical approach to deal with the dynamical rise of phase synchrony between multiple  spiking neural circuits. Making use of a mixture of mathematical techniques - mean-field theory, reduction methods, PRC measures and the framework of weakly coupled oscillators - we have been able to reduce the complexity of the problem to a single phase equation. The obtained dynamical equation reflects the contribution of cortical structure to the coordination of the macroscopic firing patterns: notably the role of the effective distance between the circuits, the native of the emergent local rhythms and the structure of the synaptic coupling across the circuits. We have shown that this level of abstraction suffices to qualitatively reproduce and explain experimentally observed oscillatory patterns. For instance, our synaptic theory allows us to clarify on the observed diversity of phase lags between multiple cortical gamma rhythms that have been proposed to play a crucial role in controlling and selecting information through anatomical pathways \cite{Fries2016}. 

Furthermore, our technique allows us to determine the directionality of cause signal transfer between multiple interacting neural circuits with emergent gamma oscillations. Using the PRC technique, we first confirmed that the signal transfer is undirected in dynamical states with full symmetry: the global PRCs were identical or just phase shifted for in-phase and anti-phase synchrony. For dynamical symmetry-broken states, where the circuits separate into a leader and a follower (also sometime called stuttering states), the global PRCs depend qualitatively on where the signal originates (e.g. in the leader) and where it propagates (e.g. to the follower). Our results show that depending on this and on the timing of the external signal perturbations, the neural activity can be either advanced or delayed. Once again, this causal functional directionality in the communication between neural circuits appears as a consequence of the system dynamics and despite a completely mirror symmetric structural connectivity and the individual network properties. We believe that these results give a causal basis for the recent statistical directed functional connectivity measures.

In the end, the series of mathematical arguments leads to a simple visualization technique - a bifurcation diagram - which compiles all the relevant information about circuit phase relationships when parameters are changed. Such graphical representation demonstrate that, in multiple delayed-coupled spiking networks, phase locking of the emergent macroscopic oscillatory rhythms are natural features that can be controlled. 
Our synaptic theory sheds new light on the long range cortical circuit interactions, and importantly, it offers a way to make strong predictions that can be tested against experimental data. For instance, one can compare the phase locking modes generated by different brain areas with distinct synaptic organization of the model. 

The formalism employed within the paper requires pyramidal neurons to work in a regime where projections across circuits are weak.
Within this parameter regime, the presented sequence of theoretical arguments are fully valid.  How our results extend to the strongly coupled regime remains a challenging topic for future studies.

Although we have restricted our study to considering
networks with homogenous synaptic weights and current-based
synaptic interaction, the mathematical strategy that served throughout this paper is adjustable and easily accepts the inclusion of conductance-based synaptic description with a certain level of synaptic heterogeneity \cite{montbrio2015,Ratas2016}. Similarly the accommodation of delay within the circuits themselves would not bring difficulty, neither for the reduction method \cite{Pazo2016}, nor for the PRC computation \cite{Kotani2012}. This could be an interesting subject of research for future works as well as the study of locking to an external periodic modulation for which the PRC offers several path of investigation \cite{Kuhn2017,Akao2018}.

All along the paper, we have only been interested to study the locking of oscillations having identical properties, however, several studies have reported coupling across different frequency band of oscillations \cite{Canolty2010}. Termed as cross-frequency coupling, the locking of brain regions with different frequencies is an open subject of research. A promising extension would then be to generalize the phase locking analysis to a network of subsequent layers made up of diversified interneuron types \cite{Jiangaac9462,Potjans2014,Bos2016}. Such a consideration would clarify the specific roles of each layer and cell types in the generation of locking and elucidate the underlying synaptic mechanism and functional roles of cross-frequency coupling observed in slow-fast oscillations \cite{Canolty2010}. To that end, one would need to study the case of interacting circuits with different intrinsic frequency which remains for us an open issue to be investigated.

\section*{Method}
\subsection*{Mean-Field}
We consider an all-to-all coupled network made up of $N$ spiking cells characterized by the quadratic integrate-and-fire (QIF) model:
\begin{equation*}
\tau \frac{d}{dt}v_j(t)=\eta_j+v_j^2(t)+I(t), 
\end{equation*}%
where $v(t)$ represents the time evolution of the membrane potential, $\tau$ is the membrane time constant, $I(t)$ is the total current, and we assume the intrinsic parameter $\eta$ being randomly distributed across the network according to a Lorentzian distribution: 
\begin{equation*}
\mathcal{L}(\eta)= \dfrac{1}{\pi} \dfrac{\Delta}{(\eta - \bar{\eta})^2 +\Delta^2}.
\end{equation*}
with $\bar \eta$ the mean value and $\Delta$ the half-width of the distribution. 
The onset of an action potential is taken into account by a discontinuous mechanism with a threshold $v_{th}$ and a reset parameter $v_r$ respectively set at plus and minus infinity \cite{Izi}.  
The population firing rate is then given by the sum of all the spikes:

\begin{equation*}
r(t)= \dfrac{1}{N}  \sum_{k=1}^{N}\sum_{f}\delta(t-t_f^k) 
\end{equation*}
where $\delta$ is the Dirac mass measure and $t_f^k$ are the firing times of the neuron numbered $k$.

In the mean-field limit, that is, when the number of cells is taken infinitely large, see  \cite{Deco2008} for instance, the system is well represented by the probability of finding the membrane potential of any randomly chosen cell at potential $v$ at time $t$ knowing the value $\eta$ of its intrinsic parameter. The dynamic of this density, which we denote $p(t,v\vert \eta )$, is given by a continuous transport equation written in the form of a conservation law:

\begin{equation}\label{PDE}
\tau   \frac{\partial}{\partial t}p(t,v\vert \eta )+ \frac{\partial}{\partial v} \mathfrak{J}(t,v\vert \eta )=0,
\end{equation}
where the total probability flux is defined as
\begin{equation*}
\mathcal{J}(t,v\vert \eta )= (\eta+v^2+I(t))p(t,v \vert \eta ).
\end{equation*}
A boundary condition, consistent with the reset mechanism of the QIF model, is imposed:

\begin{equation*}
\lim_{v \to -\infty}\mathcal{J}(t,v\vert \eta )=\lim_{v \to +\infty} \mathcal{J}(t,v\vert \eta ).
\end{equation*}
One can check easily the conservation property of the equation:

\begin{equation*}
\int_{-\infty}^{+\infty } p(t,v \vert \eta ) \, dv = \mathcal{L}(\eta) .
\end{equation*}
Importantly, the firing rate of the population $r(t)$ can be extracted from the mean-field equation, defining:

\begin{equation*}
r (t , \eta )= \lim_{v \to +\infty} \mathcal{J}(t,v\vert \eta ),
\end{equation*}
the firing rate is then given by the total probability flux crossing the threshold:
\begin{equation*}
r (t)= \lim_{v \to +\infty} \int_{-\infty}^{+\infty } \mathcal{L} (\eta) r (t, \eta )\, d\eta .
\end{equation*}

\subsection*{Reduction}
The reduction method,  see \cite{montbrio2015}, consists in assuming that the solution of the mean-field equation (\ref{PDE}) has the form of a Lorentzian distribution:

\begin{equation}\label{OAa}
p(t,v\vert \eta ) = \dfrac{1}{\pi} \dfrac{x(t,\eta)}{(v - y(t,\eta))^2 +x(t,\eta)^2}.
\end{equation}
The mean potential and the firing rate are related to the Lorentzian coefficients:

\begin{equation*}
r (t ,\eta )=\dfrac{1}{\pi} x(t, \eta),
\end{equation*} 
and 

\begin{equation*}
y(t,\eta)=\int_{-\infty}^{+\infty } v p(t,v\vert \eta )   \, dv.
\end{equation*}
Thus the mean membrane potential of the network is

\begin{equation*}
V(t)= \int_{-\infty}^{+\infty }  \mathcal{L} (\eta) y (t , \eta ) \, d \eta.
\end{equation*} 
Note  that  integrals  are  defined  via  the  Cauchy principal value, the reason being that the Lorentz distribution
only has a mean in the principal value sense.
After algebraic manipulation, see \cite{montbrio2015}, the transport equation (\ref{PDE}) reduces to the dynamical system:

\begin{equation*}
\left\lbrace
\begin{split}
&\tau  \frac{d}{dt}r = \dfrac{\Delta_e}{\pi \tau} +2  r V \\ 
&\tau  \frac{d}{dt}V = V^2 +\bar \eta  +I  - \tau^2 \pi ^2 r ^2,\\
\end{split}
\right.
\end{equation*}
Such a reduced description has the tremendous advantage to be low dimensional.

\subsection*{E-I Interaction}
Considering now a network of two interacting neural populations of excitatory cells and inhibitory cells, the system is then represented by two probability density functions, one for the excitatory population, which we denote  $p_e(t,v\vert \eta )$, and one for the inhibitory neurons, which we denote $p_i(t,v\vert \eta )$. Each density function follows a continuous transport equation similar to (\ref{PDE}).  In our case, the dynamic of the two coupled PDEs that describe the time evolution of $p_e(t,v\vert \eta )$  and $p_i(t,v\vert \eta )$ reduces to a set of differential equations. For the E-cells, we have:

\begin{equation*}
\left\lbrace
\begin{split}
&\tau_e  \frac{d}{dt}r_e = \dfrac{\Delta_e}{\pi \tau_e} +2  r_e V_e \\ 
&\tau_e  \frac{d}{dt}V_e = V_e^2 +\bar \eta_e  +I_e  - \tau_e^2 \pi ^2 r_e ^2,\\
 \end{split}\right.
\end{equation*}
and for the I-cells:

 \begin{equation*}
\left\lbrace
\begin{split}
&\tau_i  \frac{d}{dt}r_i = \dfrac{\Delta_i}{\pi \tau_i} +2  r_i V_i \\ 
&\tau_i  \frac{d}{dt}V_i = V_i^2 +\bar \eta_i  +I_i (t) - \tau_i^2\pi ^2 r_i ^2.\\
\end{split}\right.
\end{equation*}
Note that the two systems are in interaction via the expression of the currents $I_e$ and $I_i$ which include self-recurrent connections and synaptic projections, see Fig. \ref{fig01} for a schematic view. For the E-cells, the total current has the following form:

\begin{equation*}
I_e(t)= I_e^{ext}(t)+  \tau_e s_{ee}(t) - \tau_e s_{ei}(t),  
\end{equation*}%
and for the I-cells:
\begin{equation*}
I_i(t)= I_i^{ext}(t)+  \tau_i s_{ie}(t) - \tau_i s_{ii}(t),  
\end{equation*}%
here, $s_{\alpha \beta}(t)$ represents the time evolution of the synaptic current of the population $\beta$ projected on the population $\alpha$ and is given by an exponential filter of the firing activity. In the end, we get that the dynamic of the cortical network is well described by the following set of eight differential equations:

\begin{equation}\label{RedEA}
\left\lbrace
\begin{split}
&\tau_e  \frac{d}{dt}r_e = \dfrac{\Delta_e}{\pi \tau_e} +2 r_e V_e \\ 
&\tau_e  \frac{d}{dt}V_e = V_e^2 +\bar \eta_e  +I_e^{ext}+  \tau_e s_{ee} - \tau_e s_{ei}   - \tau_e^2 \pi ^2 r_e ^2\\
& \tau_s \frac{d}{dt} s_{ee}=-s_{ee} +J_{ee} r_e \\
& \tau_s \frac{d}{dt} s_{ei}=-s_{ei} +J_{ei} r_i , \\
 \end{split}\right.
\end{equation}
and
 \begin{equation}\label{RedIA}
\left\lbrace
\begin{split}
&\tau_i  \frac{d}{dt}r_i = \dfrac{\Delta_i}{\pi \tau_i} +2  r_i V_i \\ 
&\tau_i  \frac{d}{dt}V_i = V_i^2 +\bar \eta_i  + I_i^{ext}+  \tau_i s_{ie} - \tau_i s_{ii}  - \tau_i^2 \pi ^2 r_i ^2\\
&\tau_s \frac{d}{dt} s_{ie}=-s_{ie} +J_{ie} r_e \\
& \tau_s \frac{d}{dt} s_{ii}=-s_{ii} +J_{ii} r_i , \\
\end{split}\right.
\end{equation}
where $\tau_s$ is the synaptic time constant, and $J_{\alpha \beta}$ is the synaptic strength of the population $\beta$ projecting on the population $\alpha$.

\subsection*{Phase Response Curve}
The infinitesimal phase resetting curve (iPRC) is defined mathematically for infinitesimally small perturbation, and it is computed in a perfectly rigorous way via the adjoint method \cite{Brown2004}. Let us consider a general dynamical system:

\begin{equation*}
\frac{d}{dt}x(t) = F(x(t)),
\end{equation*}
where $x\in \mathcal{R}^n$. Assuming that the system admits a stable limit cycle $x_0(t)$, then if the system is perturbed by a small perturbation, the solution can be written as  

\begin{equation*}
x(t)= x_0(t)+ \epsilon p(t),
\end{equation*}
where $p(t)$ is the small deviation from the limit cycle. Up to a linearization, we get that

\begin{equation*}
\frac{d}{dt}p(t) = DF(x_0(t))\cdot p(t),
\end{equation*}
where $DF(x_0(t))$ is the time dependent Jacobian matrix. The iPRC is then defined as
\begin{equation*}
\frac{d}{dt}\left( Z(t) \cdot p(t)\right)=0,
\end{equation*}
which is equivalent to
\begin{equation*}
\begin{split}
\frac{d}{dt}\left( Z(t) \cdot p(t)\right)  	& =  \frac{d}{dt}  Z(t) \cdot p(t) + Z(t) \cdot  \frac{d}{dt} p(t)\\
							& =  \frac{d}{dt}  Z(t) \cdot p(t) + Z(t) \cdot DF(x_0(t))\cdot p(t)\\
							& =  \frac{d}{dt}  Z(t) \cdot p(t) +DF(x_0(t))^T \cdot Z(t) \cdot  p(t)\\
							& =  \left( \frac{d}{dt}  Z(t) +DF(x_0(t))^T \cdot Z(t) \right) \cdot  p(t)\\
							& = 0.\\
\end{split}
\end{equation*}
Since the last equation is valid for every perturbation $p(t)$, we get that the iPRC is solution of the adjoint equation:

\begin{equation*}
\frac{d}{dt}  Z(t) =- DF(x_0(t))^T \cdot Z(t) .
\end{equation*}
This method can be applied on the low dimensional system (\ref{RedEA})-(\ref{RedIA}) and a semi analytical expression of the iPRC  can be extracted. Assuming that

\begin{equation*} 
O(t)= \left(  r_{e_o}(t), V_{e_o}(t), s_{ee}(t), s_{ei}(t),r_{i_o}(t) ,V_{i_o}(t),s_{ie}(t), s_{ii}(t) \right), 
\end{equation*}
is a stable limit cycle of the E-I system (\ref{RedEA})-(\ref{RedIA}) of period $T$, that is, $$ O(t)= O(t+T)$$
we find that the iPRC $Z(t)$ is a periodic vector of eight components 

\begin{equation*}
Z(t)= \left(  Z_{r_e}(t), Z_{v_e}(t),  Z_{s_{ee}}(t), Z_s{_{ei}}(t), Z_{r_i}(t) ,Z_{v_i}(t) ,  Z_{s_{ie}}(t), Z_s{_{ii}}(t)  \right),
\end{equation*}
that is a solution of the adjoint equation

\begin{equation*}
-\frac{d}{dt}Z(t) = \mathcal{M}(t)^T \cdot Z(t),
\end{equation*}
where the matrix $\mathcal{M}(t) $ is given by a linearization of the E-I system (\ref{RedEA})-(\ref{RedIA}) around the limit cycle:

\begin{equation*}
\mathcal{M}(t) =
  \begin{bmatrix}
   \dfrac{2V_{e_o}(t)}{\tau_e}                    &  \dfrac{2r_{e_o}(t)}{\tau_e}   & 0          & 0         & 0                                   & 0  &0& 0   \\
    -2 \tau_e \pi^2 r_{e_o}(t)           & \dfrac{2V_{e_o}(t)}{\tau_e}  & 1 & -1 &        0                                       & 0 &0 & 0   \\
   \dfrac{ J_{ee} }{\tau_s}                                                 & 0                                       &-    \dfrac{1}{\tau_s}         &0            &0                            & 0 &0&0 \\
             0                                                      &0                                  & 0  &-    \dfrac{1}{\tau_s}  & \dfrac{ J_{ei} }{\tau_s}                              & 0 &0&0 \\
     0                                                         &  0                                           &  0   & 0 &     \dfrac{2V_{i_o}(t)}{\tau_i}    &   \dfrac{2r_{i_o}(t)}{\tau_i} & 0 &0 \\
    0                                     &  0                                                  &     0& 0 &     -2 \tau_i \pi^2 r_{i_o}(t)    &  \dfrac{2V_{i_o}(t)}{\tau_i} &  1  & - 1  \\
           \dfrac{ J_{ie} }{\tau_s}                                                                                       &     0        &0&0   & 0  &0            & -    \dfrac{1}{\tau_s}                                   &0   \\
              0                                                      & 0                                  &0 &0&    \dfrac{ J_{ii} }{\tau_s}                             & 0 &0&-    \dfrac{1}{\tau_s}      \\
  \end{bmatrix}.
\end{equation*}
The iPRC $Z(t)$ is given by the unique periodic solution that satisfies the normalization condition $$Z(t)\cdot \dot O(t) = 2\pi / T.$$

\begin{figure}[t!]
\begin{center}  
      \includegraphics[width=\textwidth]{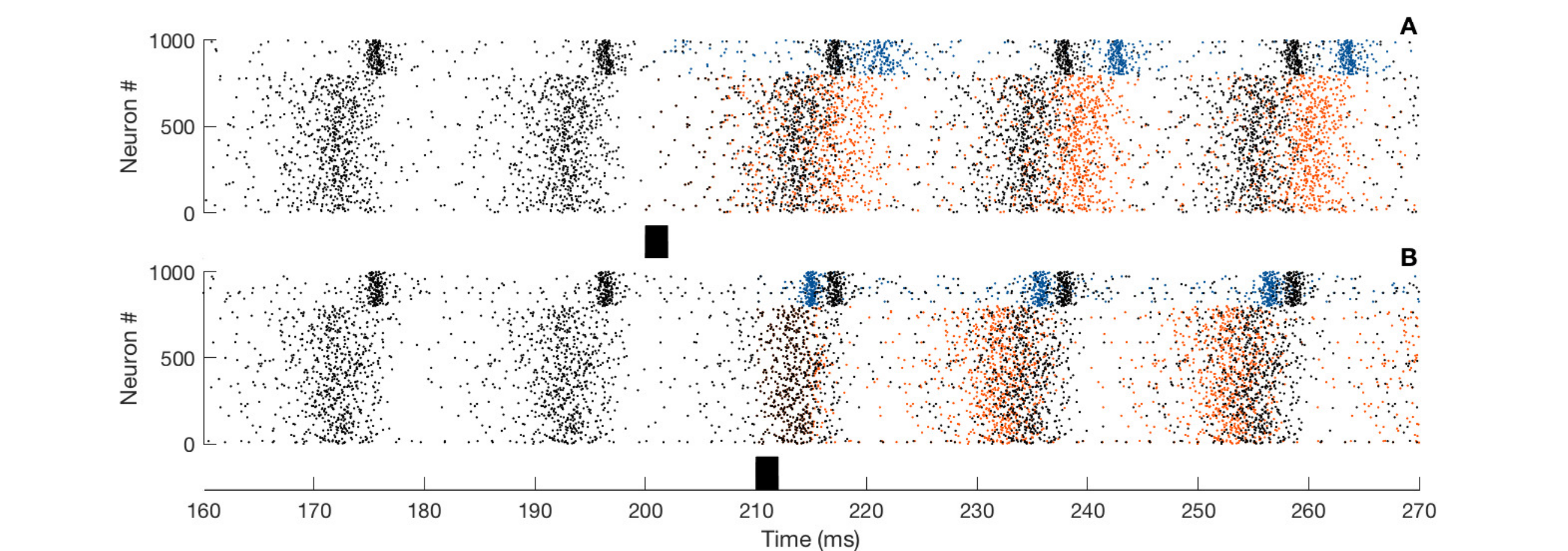}   
   \caption{Phase shifting.  A-B) Spiking activity obtained from simulations of the full spiking network.  The black dots illustrate the ongoing activity and the colored dots (blue for the I-cells and red for the E-cells) the activity of the perturbed network. Lower panels: Illustration of the stimulus onset. Perturbations are made on  the I-cells. The network parameters are the same as in Fig. \ref{fig04} with $I_e^{ext}=10$, direct perturbations are made with a square wave current pulse (amplitude $10$, duration $0.5$).  }
    \label{fig06}
      \end{center}
\end{figure}

The iPRC can be compared with a direct method which consist in presenting perturbation to the network. Depending on the phase onset of the perturbation, the network activity is going to shift.
Raster plots from numerical simulations of the full network (Fig. \ref{fig06}) illustrate the shift. Here the black dots correspond to the unperturbed network, whereas the colored dots to the perturbed circuit. Before the stimulus onset, the two rasters overlap perfectly. After the stimulus presentation, spikes of the perturbed network are shifted: either delayed (Fig. \ref{fig06}A) or advanced (Fig. \ref{fig06}B) depending on the onset phase of the perturbation. In the Result section, the two approches - direct perturbation and the adjoint method - are compared for the PING interaction in Fig. \ref{fig04} and the ING interaction in Fig. \ref{fig05}.

\subsection*{Bidirectionally Coupled Networks}
Considering now two bidirectionally delayed coupled networks where the coupling is made via long projections of the pyramidal cells from one network to another, the whole system reduces to a set of sixteen differential equations. For the first network, we have

\begin{equation}\label{RedE1}
\left\lbrace
\begin{split}
&\tau_e  \frac{d}{dt}r_{e_1} = \dfrac{\Delta_e}{\pi \tau_e} +2  r_{e_1} V_{e_1} \\ 
&\tau_e  \frac{d}{dt}V_{e_1} = V_{e_1}^2 +\bar \eta_e  + +I_e^{ext}+  \tau_e s_{ee_1} - \tau_e s_{ei_1}   - \tau_e^2 \pi ^2 r_{e_1} ^2\\
&\tau_s \frac{d}{dt} s_{ee_1}=-s_{ee_1} +J_{ee} r_{e_1} + G_{ee}r_{e_2}(t-d)  \\
&\tau_s \frac{d}{dt} s_{ei_1}=-s_{ei_1} +J_{ei} r_{i_1} ,  \\
 \end{split}\right.
\end{equation}
and

 \begin{equation}\label{RedI1}
\left\lbrace
\begin{split}
&\tau_i  \frac{d}{dt}r_{i_1} = \dfrac{\Delta_i}{\pi \tau_i} +2  r_{i_1} V_{i_1} \\ 
&\tau_i  \frac{d}{dt}V_{i_1} = V_{i_1}^2 +\bar \eta_i  + I_i^{ext}+  \tau_i s_{ie_1} - \tau_i s_{ii_1}  -\tau_i^2 \pi ^2 r_{i_1} ^2\\
&\tau_s \frac{d}{dt} s_{ie_1}=-s_{ie_1} +J_{ie} r_{e_1} + G_{ie}r_{e_2}(t-d)   \\
&\tau_s \frac{d}{dt} s_{ii_1}=-s_{ii_1} +J_{ii} r_{i_1},  \\
\end{split}\right.
\end{equation}
and for the second network:

\begin{equation}\label{RedE2}
\left\lbrace
\begin{split}
&\tau_e  \frac{d}{dt}r_{e_2} = \dfrac{\Delta_e}{\pi \tau_e} +2  r_{e_2} V_{e_2} \\ 
&\tau_e  \frac{d}{dt}V_{e_2} = V_{e_2}^2 +\bar \eta_e  + +I_e^{ext}+  \tau_e s_{ee_2} - \tau_e s_{ei_2}   -\tau_i^2 \pi ^2 r_{e_2} ^2\\
&\tau_s \frac{d}{dt} s_{ee_2}=-s_{ee_2} +J_{ee} r_{e_2} + G_{ee}r_{e_1}(t-d)  \\
&\tau_s \frac{d}{dt} s_{ei_2}=-s_{ei_2} +J_{ei} r_{i_2} ,  \\
 \end{split}\right.
\end{equation}
and

 \begin{equation}\label{RedI2}
\left\lbrace
\begin{split}
&\tau_i  \frac{d}{dt}r_{i_2} = \dfrac{\Delta_i}{\pi \tau_i} +2  r_{i_2}  V_{i_2}  \\ 
&\tau_i  \frac{d}{dt}V_{i_2} = V_{i_2} ^2 +\bar \eta_i  + I_i^{ext}+  \tau_i s_{ie_2} - \tau_i s_{ii_2}  - \tau_i^2  \pi ^2 r_{i_2}  ^2\\
&\tau_s \frac{d}{dt} s_{ie_2}=-s_{ie_2} +J_{ie} r_{e_2}  + G_{ie}r_{e_1}(t-d)  \\
&\tau_s \frac{d}{dt} s_{ii_2}=-s_{ii_2} +J_{ii} r_{i_2} ,  \\
\end{split}\right.
\end{equation}
Note the presence of long range projections between circuits, see Fig. \ref{fig05} for a schematic view. Here $G_{\alpha \beta}$ denotes the connectivity strength of the population $\beta$ of one network onto the population $\alpha$ of the other circuit, and the parameter $d$ is the conduction delay between the two networks.

\subsection*{Phase Equation}
Assuming that the two networks are oscillating and placing our study within the framework of weakly coupled oscillators, that is, if we assume that $$G_{\alpha \beta}<<1,$$  we can reduce the bidirectionally delayed-coupled neural circuits description (\ref{RedE1})-(\ref{RedI1})-(\ref{RedE2})-(\ref{RedI2}) to a single phase equation:

\begin{equation*}
\dfrac{d}{dt} \theta(t) = G(\theta(t)).
\end{equation*}
Here $\theta(t)$ is the phase difference (or phase lag) between the circuits and the $G$-function is the odd part of the shifted interaction function (the $H$-function), see \cite{Ermentrout1097} for instance:

\begin{equation*} 
 G(\theta) = H (\theta - d) -H(-\theta - d)  ,
\end{equation*}
with $d$, the time delay between the two circuits. In our case, the interaction function is mathematically described as

\begin{equation*}
\begin{split}
H(\theta) = & \dfrac{G_{ee}}{T}\int_0^T Z_{s_{ee}}(s)r_e(s-\theta) \, ds  \\
& + \dfrac{G_{ie}}{T}\int_0^T Z_{s_{ie}}(s)r_e(s-\theta) \, ds ,
\end{split}
\end{equation*} 
where $T$ is the oscillation period.
Note the involvement of the synaptic component of the PRC $Z_s(t)$ and the firing rate of the E-cells $r_e(t)$ all along the oscillatory cycle in the expression of the $G$-function.

\section*{Acknowledgments}
This  study  was  funded  by  CNRS,  INSERM,  and  partial
support from LABEX ANR-10-LABX-0087 IEC and IDEX
ANR-10-IDEX-0001-02  PSL*.  The  study  received  support
from the Russian Science Foundation grant (Contract No. 17-
11-01273).


\end{document}